\documentclass[12pt]{article}
\usepackage{graphicx}
\usepackage{amsmath,amssymb}
\usepackage{scicite}
\usepackage{xcolor}
\usepackage{lineno}
\usepackage{siunitx}
\usepackage{todonotes}
\usepackage{hyperref}
\hypersetup{
    colorlinks=true,
    linkcolor=blue,
    filecolor=magenta,      
    urlcolor=blue,
    pdftitle={Evidence for neutrino emission from the nearby active
galaxy NGC 1068},
    pdfpagemode=FullScreen,
    }

\usepackage{times}

\newenvironment{sciabstract}{%
\begin{quote} \bf}
{\end{quote}}

\newcounter{lastnote}

\title{Evidence for neutrino emission from the nearby active galaxy NGC\,1068} 

\author
{IceCube Collaboration\footnote{The full list of collaboration members and their affiliations is included in the supplementary material}\\
\\
\\
\normalsize{$^\ast$E-mail:  analysis@icecube.wisc.edu.}
}

\date{}

\begin{document} 
\baselineskip24pt
\maketitle 

\begin{sciabstract}
  We report three searches for high energy neutrino emission from astrophysical objects using data recorded with IceCube between 2011 and 2020. Improvements over previous work include new neutrino reconstruction and data calibration methods. In one search, the positions of 110 a priori selected gamma-ray sources were analyzed individually for a possible surplus of neutrinos over atmospheric and cosmic background expectations. We found an excess of $79_{-20}^{+22}$ neutrinos associated with the nearby active galaxy NGC\,1068 at a significance of 4.2$\,\sigma$. 
 The excess, which is spatially consistent with the direction of the strongest clustering of neutrinos in the Northern Sky, is interpreted as direct evidence of TeV neutrino emission from a nearby active galaxy. The inferred flux exceeds the potential TeV gamma-ray flux by at least one order of magnitude.
\end{sciabstract}

\section*{High energy cosmic neutrinos}
The observation of the highest-energy cosmic protons and nuclei \cite{Abraham:2004dt, ThePierreAuger:2015rma,AbuZayyad:2012kk} has revealed the existence of powerful cosmic particle accelerators but not their nature and location in the universe. 
Due to the presence of interstellar magnetic fields, charged cosmic particles change direction during their propagation to Earth. Thus, the quest to identify cosmic accelerators relies on high-energy  photons and neutrinos. Both travel along straight paths and are produced wherever cosmic rays interact with ambient matter, mainly protons, or photons in or near the acceleration sites. Depending on the environment in which these interactions occur, gamma rays can rapidly lose energy via several processes, such as pair-production through the interaction with lower energy photons. Similarly, above TeV energies gamma rays are strongly absorbed over cosmological distances through interactions with the extragalactic background light and the cosmic microwave background \cite{DeAngelis:2013jna}.  Neutrinos are not affected by absorption during their cosmic journey, and thus are excellent probes of the most extreme cosmic accelerators.
Being among the most energetic and enigmatic astrophysical sources in the universe, active galaxies, i.e. galaxies that host an Active Galactic Nucleus (AGN) \cite{Padovani:2017zpf}, have long been considered as candidate neutrino emitters \cite{1977Berezinsky, 1979ApJ...232..106E, 1981MNRAS.194....3B, 1989A&A...221..211M, Halzen:1997hw}. The accretion flow of matter into a central supermassive black hole (SMBH) powers the AGN, which is typically surrounded by an obscuring, dusty torus, and thus the observable characteristics of the AGN depend on the viewing angle. For example, Seyfert II galaxies \cite{Seyfert:1943} are assumed to be viewed edge on with the line of sight passing directly through the obscuring torus \cite{Antonucci:1993}. In some cases, the AGN can launch a strong, narrow jet of accelerated plasma. If such a jet is oriented close to the line of sight, the AGN is also called a blazar \cite{Urry:1995mg}.

The IceCube Neutrino Observatory \cite{Aartsen:2016nxy}, based at the Amundsen-Scott South Pole Station in Antarctica, is currently the world's largest detector for neutrinos. Its construction was completed on December 18, 2010. The telescope uses one cubic kilometer of the optically transparent glacial ice as a detection medium to measure the Cherenkov light emitted by relativistic charged particles. These secondary particles are created when neutrinos interact in or near the instrument. In total, 5160 digital optical modules (DOMs) are installed on 86 vertical cables (``strings''), spaced 125\,meters apart, thus forming a three-dimensional array in the ice. Each DOM records the number of induced photo-electrons, or charges, as a function of time. In 2013, IceCube announced the detection of high-energy neutrinos of extra-terrestrial origin \cite{Aartsen:2013jdh}.
The measured flux of these astrophysical neutrinos appears largely isotropic, equally distributed among flavors, and can be described by a single power-law energy distribution that extends from $\sim$10\,TeV to PeV energies \cite{Aartsen:2016xlq, Aartsen:2020aqd, Abbasi:2020zmr, IceCube:2021rpz, PhysRevD.104.022002}, but its origin remains largely unresolved.
To date the most compelling evidence of a source of high-energy cosmic neutrinos was reported following the space and time coincidence of a high-energy IceCube neutrino \cite{Aartsen:2016lmt} with the gamma-ray flaring blazar TXS\,0506+056 \cite{IceCube:2018dnn, IceCube:2018cha, Padovani:2019xcv}. 
TXS\,0506+056 has recently been shown to host a standard accretion disk and a dusty torus, which has the peculiarity to be also an emitter of high-energy radiation and, possibly, cosmic rays \cite{Padovani:2019xcv}.
Additionally, a recent IceCube study \cite{Aartsen:2019fau} suggests a correlation of IceCube neutrinos with a catalog composed of 110 known gamma-ray emitters at a significance of 3.3$\,\sigma$. The main contributors are the active galaxy NGC\,1068 and the blazars TXS\,0506+056, PKS\,1424+240 and GB6\,J1542+6129. The significance of the neutrino excess from the direction of NGC\,1068 was reported as 2.9$\,\sigma$ \cite{Aartsen:2019fau}.

Here, we present an improved search for neutrino point sources based on new data processing, data calibration, and event reconstruction methods that lead to a more precise neutrino event characterization and thereby to better signal and background discrimination, as well as a more accurate estimate of the neutrino fluxes from potential neutrino sources.

\section*{Searching the Northern Sky for point-like neutrino emission}
In this paper, we present analyses of data collected with IceCube between May 13, 2011, and May 29, 2020. This period covers all the data recorded with the full 86-string detector configuration. In contrast, a recently reported result from searches for cosmic neutrino sources\cite{Aartsen:2019fau} also included the partial detector configurations with fewer strings, going back to 2008, and stopping in spring 2018. In this analysis, we only use the full detector data as our methods depend on uniformly processed data. We have verified that the high-statistics Monte Carlo dataset agrees very well with the observed data.
The IceCube dataset \cite{Aartsen:2018ywr}, constructed using selection criteria presented in \cite{Aartsen:2016xlq}, has been reprocessed uniformly, removing the data sample fragmentation in earlier results, aligning different data-taking conditions and calibrations, and improving event reconstructions and resolutions \cite{note:Suppl-Mat}. Specifically, for this reprocessing we consistently apply the most accurate version of the directional track reconstruction method (\textit{SplineReco} algorithm \cite{AMANDA:2003vtt, IceCube:2021oqo, note:Suppl-Mat}) to all events recorded throughout the entire data taking period \cite{note:Suppl-Mat}. Here, we use the latest calibration information that increase the accuracy in the extraction of the charges at each DOM and in the corresponding arrival times of Cherenkov photons, thus leading to small changes in the reconstructed event energies, and in some cases also to the reconstructed event directions \cite{note:Suppl-Mat}. To ensure a uniform detector response, DOMs of the DeepCore sub-array, commonly used to study $\lesssim 100\,\mathrm{GeV}$ neutrinos, are excluded \cite{Aartsen:2016xlq}.
The resulting dataset, optimised for $\nu_\mu$-induced track-like events, has a total livetime of 3186 days.
The searches presented in this work focus on the Northern Sky ($\delta\geq-3^\circ$), where IceCube sensitivity to individual astronomical objects is maximal. IceCube uses the Earth as a passive cosmic muon shield and as target material for neutrinos. Hence, by selecting upward-going events we efficiently reject the atmospheric muon background, which contributes less than $0.3\%$ to the final event sample \cite{Aartsen:2016xlq}.
A total of around 670,000 neutrino-induced muon tracks pass the final event selection criteria \cite{Aartsen:2016xlq}. However, only a small fraction of these events originate from neutrinos produced in astrophysical sources. The vast majority arises from the decay of mesons which are produced in the interaction of cosmic radiation with nuclei of the Earth's atmosphere. 
To discriminate neutrinos that originate from individual astrophysical sources from the background of atmospheric and diffuse astrophysical neutrinos, we employ the maximum-likelihood method and likelihood ratio hypothesis testing based on the estimated energy, direction, and angular uncertainty of each event \cite{note:Suppl-Mat}. The median angular resolution of the single neutrino arrival direction, composed of reconstruction error and stochasticity of kinematic processes, is 1.2$^{\circ}$ at 1\,TeV, 0.4$^{\circ}$ at 100\,TeV, and 0.3$^{\circ}$ at 1\,PeV. For the source emission, we assume in the tests a generic power-law energy spectrum, $\Phi_{\nu_{\mu}+\bar{\nu}_{\mu}}(E_{\nu}) = \Phi_0 \cdot (E_{\nu}/E_0)^{-\gamma}$, with $E_0=1\,$TeV and the spectral index $\gamma$ and the flux normalization $\Phi_0$  as free parameters \cite{note:Suppl-Mat}.
This corresponds to two correlated model parameters which are expressed as a pair ($\mu_{ns}$, $\gamma$), where $\mu_{ns}$ is the mean number of astrophysical neutrino events associated with a given point in the sky. Using the declination-dependent effective area of the detector and the spectral index $\gamma$, $\mu_{ns}$ can be directly converted to $\Phi_0$. Hence, the tuple of $\mu_{ns}$ and $\gamma$ fully determines the flux of muon neutrinos $\Phi_{\nu_{\mu}+\bar{\nu}_{\mu}}$ at any given energy.

Three different searches are presented here. The first search consists of three discrete scans of the Northern Sky to identify the location of the most significant excess of high-energy neutrino events using three different hypotheses on the spectral index: $\gamma$ as a free parameter and $\gamma$  fixed to 2.0 and 2.5. The other two tests are done on a list of 110 astronomical objects, all located in the Northern Hemisphere: one is the search for the most significant candidate neutrino source in the list, the second consists of a binomial test to evaluate the significance of observing an excess of $k$ sources with local p-value below or equal to a certain threshold, with $k$ going from 1 to 110. Similar to the sky scan, the binomial test is also repeated under the three spectral index hypotheses. All these searches are performed in the Northern Hemisphere from declination $\delta=-3^\circ$ to $\delta=81^\circ$. Higher declinations are excluded because low energy events tend to line-up completely with the strings of IceCube, complicating the numerical modeling of the signal and background in the likelihood. The resulting loss of sky coverage is less than $1\%$.

All analysis methods, including the selection of the hypotheses to be tested, were formulated \textit{a priori}. The analysis performance was evaluated using simulations, as well as randomized experimental data.
The local p-values are determined as the fraction of simulated background-only experiments, yielding a test statistic greater than or equal to the one obtained from the actual experimental data. The global p-value is determined from the smallest local p-value after correcting for multiple testing \cite{note:Suppl-Mat}. It is used to assess the evidence the data provide against the background-only assumption, i.e., that the data consist purely of atmospheric and isotropic cosmic neutrinos.

The discrete scan of the sky maximizes the signal and background likelihood function at each point on a $\sim\left(0.2^{\circ} \times 0.2^{\circ}\right)$ grid, thus obtaining the local value of the test statistic, the best-fit value of the mean number of astrophysical neutrino events ($\hat{\mu}_{ns}$), and the spectral index ($\hat{\gamma}$). The scan is repeated with fixed spectral indices $\gamma$\,=\,2.0 and $\gamma$\,=\,2.5 in order to reduce the impact of background fluctuations which arise from the conventional atmospheric neutrinos in the dataset. The conventional atmospheric spectrum follows a power law with $\gamma \approx 3.7$ and hence those fluctuations typically yield soft spectral indices with $\gamma \gtrsim 3$. Fixing the spectral index therefore increases the sensitivity to astrophysical sources with harder spectra. For the 20 most significant hot-spots in each of the three searches, the vicinity is scanned with increased resolution of $\sim\left(0.03^{\circ} \times 0.03^{\circ}\right)$ on a square grid of total size 1.5$^{\circ}$ $\times$ 1.5$^{\circ}$. Overall, the hottest spot is identified in the first scan with free spectral index at right ascension 40.69$^{\circ}$ and declination 0.09$^{\circ}$ (J2000) with $\hat{\gamma}=3.2$ and $\hat{\mu}_{ns}=81$.

\begin{figure}
\centering
\includegraphics[width=\textwidth]{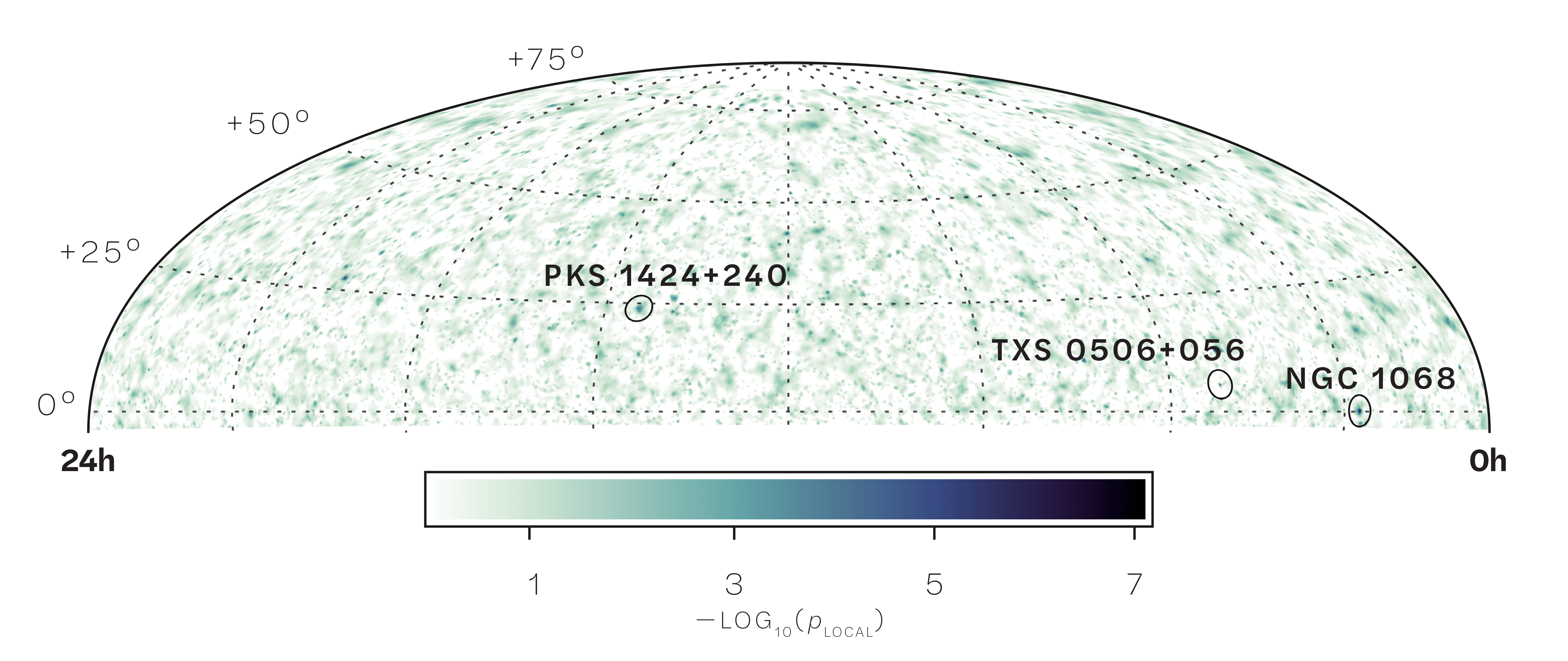}
\caption[]{{\bf Skymap of the scan for point sources in the Northern Hemisphere}. The color scale represents the local p-value obtained from the maximum likelihood analysis evaluated (with the spectral index as free fit parameter) at each location in the sky, shown in Equatorial coordinates with Hammer-Aitoff projection.
The black circles indicate the three most significant objects in the source list search. The circle of NGC\,1068 also coincides with the overall hottest spot in the Northern Sky.}
\label{fig:skymap}
\end{figure}

\begin{figure}[t]
\centering
\includegraphics[width=0.49\textwidth]{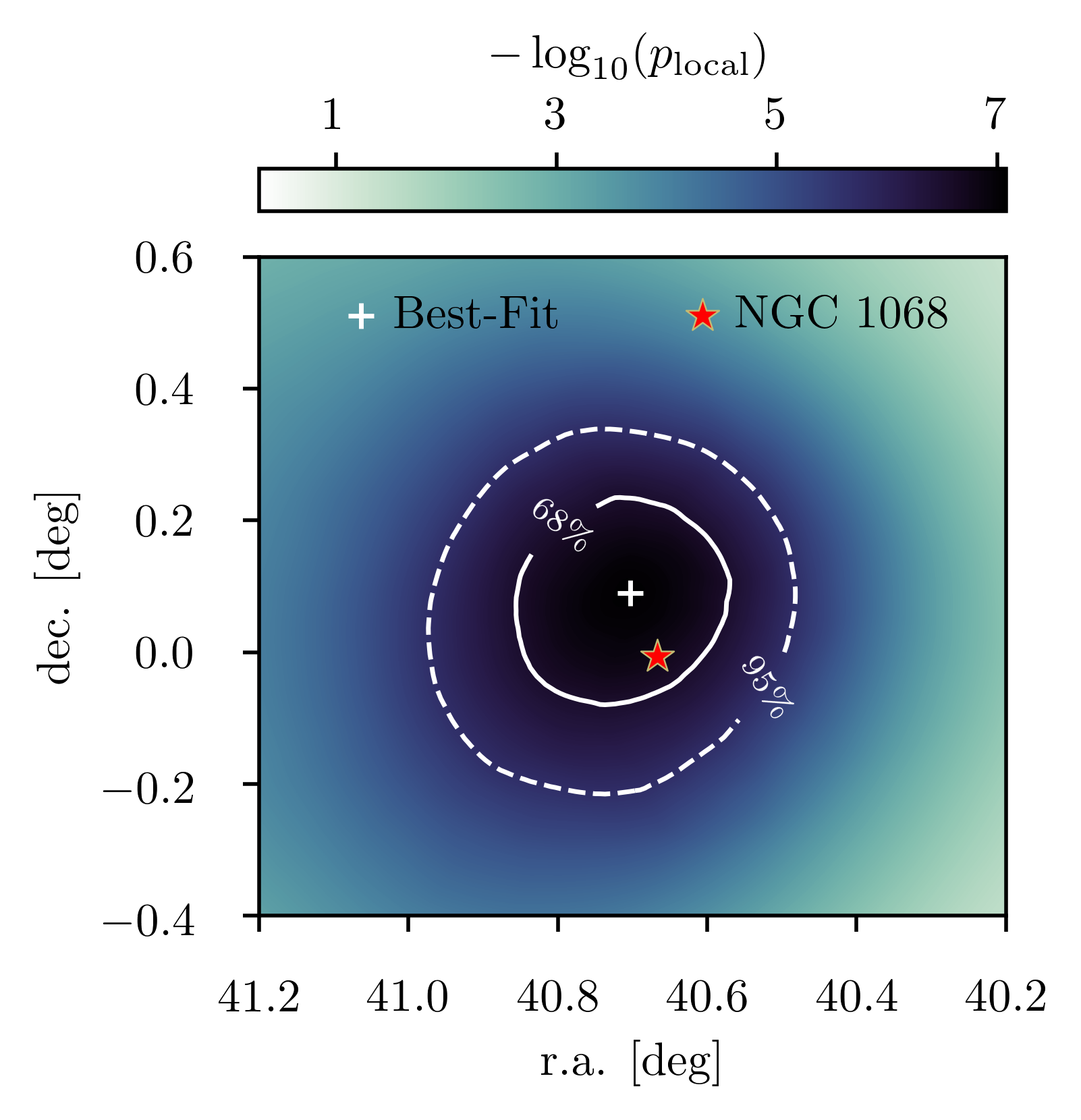}
\hfill
\includegraphics[width=0.49\textwidth]{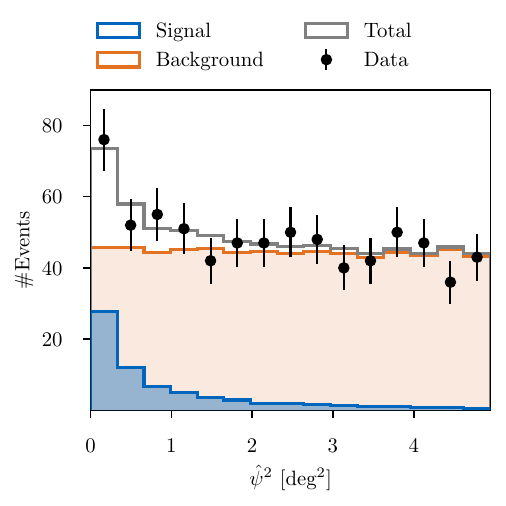}
\caption[Sky region around the hottest spot]{{\bf The sky region around the most significant spot in the Northern Hemisphere and NGC\,1068}. The left plot shows  a fine scan of the region around the hottest spot. The spot itself is marked by a yellow cross and the red star shows the position of NGC\,1068. In addition, the solid and dashed contours show the 68\% (solid) and 95\% (dashed) confidence regions of the hot spot localization. The right plot shows the distribution of the squared angular distance between NGC\,1068 and the reconstructed event direction. From Monte Carlo we estimate the background (orange) and the signal (blue) assuming the best-fit spectrum at the position of NGC\,1068. The superposition of both components is shown in gray and provides an excellent match to the data (black). Note that this representation of the result neglects all the information on the energy and angular uncertainty of the events that is used in the unbinned maximum likelihood approach.}
\label{fig:hotspot}
\end{figure}

The full skymap is shown in Figure \ref{fig:skymap}. At the location of the hottest spot, the local p-value of 5 $\times$ 10$^{-8}$ corresponds to a local significance of 5.3$\,\sigma$. After including a penalty for the multiple tests performed, i.e. asking how likely it is to observe an equal or larger significance when scanning many independent positions in the sky under the three spectral index hypotheses, the global p-value corresponds \cite{note:Suppl-Mat} to a significance of 2.0$\,\sigma$ and therefore is not significant when the entire Northern Sky is scanned without additional prior information. A high-resolution scan around the best-fit position of the hottest spot is shown in Fig.\,\ref{fig:hotspot}.

As part of the various inspections to be carried out \textit{a posteriori}, we also searched for astrophysical counterparts in close proximity with the direction of the five locally most significant spots in each of the three skymaps (reported in Tab.\,2 \cite{note:Suppl-Mat}). We note that the nearby Seyfert I galaxy NGC 4151 \cite{Seyfert:1943} is located at $\sim$0.18 degrees distance from the fourth-hottest spot in the map obtained with $\gamma$=2.5. Because possible neutrino emission from NGC 4151 is not one of the hypotheses that were formulated for this work, we cannot estimate a global p-value for this coincidence.

  Searching the entire Northern Hemisphere entails a strong penalty due to testing multiple locations. A standard way to mitigate this trial factor is to reduce the search to a list of \textit{a priori} selected positions in the sky based, for example, on known gamma-ray emission  \cite{Gaisser:2016uoy}. Such a procedure has been followed by IceCube since the first analysis \cite{IceCube:2009hni}, and revised in 2018 using recent multi-wavelength information \cite{Aartsen:2019fau}. No indication of neutrino emission from NGC\,1068 was noticeable at the time \cite{Aartsen:2016oji} when the selection method of candidate sources was updated \cite{Aartsen:2019fau}. To avoid confirmation bias the procedure from \cite{Aartsen:2019fau} was kept to select a total of 110 objects within the declination range of this analysis, $-3^{\circ} \leq \delta \leq 81^{\circ}$. The number of 110 objects was chosen such that a local $5\,\sigma$ detection will yield a final result above $4\,\sigma$ after accounting for trials. 
To select the 110 astronomical objects, we consider the gamma-ray flux above 1\,{GeV} of the {\it Fermi}-LAT sources in the 4FGL-DR2 catalog \cite{Fermi-LAT:2019yla} weighted by the IceCube sensitivity. The final catalog contains 59 BL Lacs, 34 FSRQs, 2 blazars of uncertain type, 5 AGNs, 9 galaxies, and one Galactic source from the TeVCAT catalog \cite{2008ICRC....3.1341W}, see also \cite{note:Suppl-Mat}. We do not further impose or require any relationship between the observed gamma-ray flux and the hypothesized neutrino flux when testing the selected objects for an association to neutrinos\cite{note:Suppl-Mat}.

Among the 110 astronomical objects tested, NGC\,1068 is the most significant with a local p-value of $1 \times 10^{-7}$ ($5.2\,\sigma$) and with best-fit values of spectral index $\hat{\gamma}$\,=\,3.2 and mean number of signal events $\hat{\mu}_{ns}$\,=\,79. NGC\,1068 is contained within the 68\% confidence region around the hottest spot in the Northern Sky, located at a distance of 0.11$^{\circ}$. Thus, the hottest spot location is consistent with possible neutrino emission from NGC\,1068 (Fig.\,\ref{fig:hotspot}). After correcting for having tested the 110 sources in the catalog, the global p-value for NGC\,1068 is $1.1 \times 10^{-5}$, corresponding to a significance of 4.2$\,\sigma$.\\
To investigate the source catalog for an excess of $k$ sources with local p-value below or equal to a certain threshold, we also perform a binomial test \cite{note:Suppl-Mat}, as defined \textit{a priori}. By scanning the p-value threshold, we find the smallest background probability for an excess of $k=3$ sources, tested under the free spectral index hypothesis, with local p-values smaller or equal to $4.6\times 10^{-6}$. This results in a local significance of 3.7\,$\sigma$, a small increase with respect to what was reported in \cite{Aartsen:2019fau} that is independent of the increase of the significance at the location of NGC\,1068. After correcting for having tested three different spectral index hypotheses, we obtain a final post-trial significance of 3.4$\,\sigma$ for the binomial test. Besides NGC\,1068, the other two objects contributing to the excess are the blazars PKS\,1424+240 and TXS\,0506+056, for which we find potential neutrino emission with local significance of 3.7$\,\sigma$ and 3.5$\,\sigma$, respectively. We emphasize that the significance of TXS\,0506+056 reported here relates to a time-integrated signal, whereas previous analyses have provided evidence for transient emission \cite{IceCube:2018dnn, IceCube:2018cha, Padovani:2018acg}. Such emission has not been tested in this work, which is instead optimized to measure possible steady neutrino fluxes over the whole period of the data taking. Compared to the methods presented here, a search for transient emission has a lower effective background. The total number of contributing candidate sources ($k=3$) in the binomal test has been reduced by one compared to \cite{Aartsen:2019fau}. While the local significance of PKS\,1424+240 increased from $3.0\,\sigma$ to $3.7\,\sigma$, the local significance of GB6 J1542+61 decreased from $2.9\,\sigma$ to $2.2\,\sigma$, falling now below the threshold corresponding to the best binomial p-value. The results of all three searches are summarized in Tab.\,\ref{Tab:final-results}. Additional details can be found in \cite{note:Suppl-Mat}.

\begin{table}[t]
\begin{center}
\begin{tabular}{ l c c }
\hline \hline
 Test type & Pre-trial p-value (p$_{local}$) & Post-trial p-value (p$_{global}$) \\ 
 \hline
 Northern Hemisphere scan & $5.0 \times 10^{-8}$ (5.3$\,\sigma$) & $2.2 \times 10^{-2}$  (2.0$\,\sigma$)\\ 
 List of candidate sources, single test & $1.0\times 10^{-7}$ (5.2$\,\sigma$) & $1.1\times 10^{-5}$ ($4.2\,\sigma$) \\  
List of candidate sources, binomial test & $4.6\times 10^{-6}$ (4.4$\,\sigma$) & $3.4\times 10^{-4}$ (3.4$\,\sigma$)  \\
 \hline
\end{tabular}
\caption[Final results]{{\bf  Summary of final p-values.} For each of the three tests performed in this work, we report the local and global best p-values. }
\label{Tab:final-results}
\end{center}
\end{table}

\section*{Neutrino emission from the direction of NGC\,1068}
A high-resolution scan around the most significant spot in the Northern Hemisphere is shown in Fig.\,\ref{fig:hotspot} (left), with NGC\,1068 located inside the 68\% confidence region. The excess at the position of NGC\,1068 is produced by $\hat{\mu}_{ns}$=79$^{+22}_{-20}$ events above the expectation from the atmospheric and diffuse astrophysical neutrino backgrounds. The systematic error on $\hat{\mu}_{ns}$ is around $2$ events \cite{note:Suppl-Mat}. Among the 79 most contributing events, 63 were recorded during the time period that overlaps with the previous analysis \cite{Aartsen:2019fau} and all of them contributed to the previous result. Fig.\,\ref{fig:hotspot} shows the distribution of all events in the vicinity of NGC\,1068, as well as the confidence contours for the source location obtained in this work. The measured spectral index is $\hat{\gamma}=3.2^{+0.2}_{-0.2}$ with an estimated systematic uncertainty of $0.07$ \cite{note:Suppl-Mat} and is thus consistent with the previous result reported in \cite{Aartsen:2019fau}. Systematic uncertainties arise mainly from the modeling of the photon propagation in the glacial ice, e.g. scattering and absorption, as well as the efficiency with which they are detected by the IceCube optical modules. Systematic uncertainties are sub-dominant to statistical ones for directional track reconstructions \cite{note:Suppl-Mat}, but have a non-negligible effect on the energy reconstructions, as the observed light yield is altered.
The impact of these systematic uncertainties on the measured neutrino spectrum is estimated by analyzing simulated data, assuming a source with flux equal to the one measured for NGC\,1068 but varying ice and detector properties correspondingly \cite{note:Suppl-Mat}.

\begin{figure}[t]
   \centering
   \includegraphics[width=0.85\textwidth]{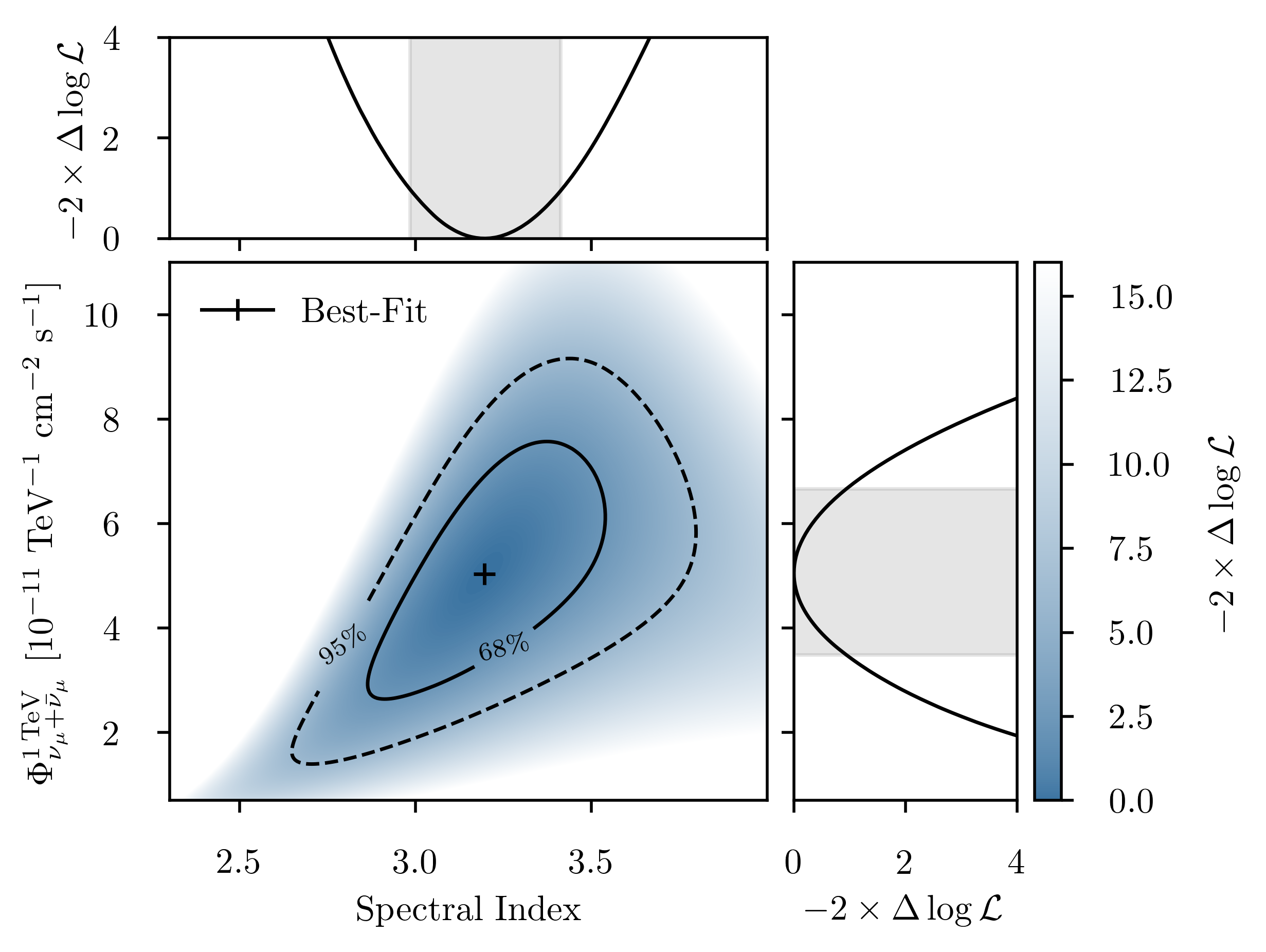}
    \caption[Profile likelihood of the NGC\,1068 neutrino spectrum]{{\bf Profile likelihood scan for the flux parameters of NGC\,1068}. The cross shows the best-fit value, solid and dashed lines represent 68\% and 95\% confidence levels derived from Wilks' Theorem, respectively. The side panels show the corresponding one-dimensional profile likelihoods. All contours include only statistical uncertainties.}
    \label{fig:profile}
\end{figure}

A detailed characterization of the source spectrum is shown in Fig.\,\ref{fig:profile}, where we report the likelihood as a function of the model parameters ($\Phi_{0}$, $\gamma$) evaluated at the coordinates of NGC\,1068. The conversion of $\hat{\mu}_{ns}$ to the flux $\Phi_{0}$ also accounts for the contribution from tau neutrino interactions which produce muons in the final state assuming an equal neutrino flavor ratio. The best-fit flux averaged over the data-taking period at a neutrino energy of 1 TeV is $\Phi_{\nu_{\mu}+\bar{\nu}_{\mu}}^{1\,\si{TeV}}$\,=\,$(5.0 \pm 1.5_{stat} \pm 0.6_{sys}) \times 10^{-11}\,\si{TeV^{-1}.cm^{-2}.s^{-1}}$. Here, the systematic uncertainty is estimated by varying the flux normalization under different ice and detector properties such that we reproduce the experimental result of $\hat{\gamma}$ and $\hat{\mu}_{ns}$ in the median case.

The unbroken power-law assumption used as spectral model in the likelihood is defined over the entire energy range of the dataset.
However, the main contribution to the excess and thus the measured spectral index and flux normalization comes from neutrinos in an energy range from 1.5\,TeV to 15\,TeV.
This energy range is defined as the range that contributes 68\% to the total test statistic value of the measurement. Outside of this energy range, the data does not strongly constrain the inferred flux properties.
Our results significantly strengthen the suggestive results reported in \cite{Aartsen:2019fau}, where this source was also identified as the strongest association, albeit at reduced significance (2.9$\,\sigma$) and larger separation from the hottest spot in the Northern sky (0.35$^{\circ}$).

The increased significance of the neutrino hotspot in spatial coincidence with NGC\,1068 can be traced to several (sometimes small) improvements applied to the entire sample before unblinding the data. The principle contributions come from 1) the new data processing, 2) the new calibration methods, and 3) the improved analysis methods and tools including a significantly more accurate characterization of IceCube point spread function and better estimation of both the event energy and angular uncertainty \cite{note:Suppl-Mat}.

For instance, a reduced significance of $3.8\,\sigma$ would have been obtained had we analyzed this sample of events based on the same up-to-date data calibration methods and associated event directions, but without these new methods. This difference is expected and can be reproduced in Monte Carlo studies, see \cite{note:Suppl-Mat}.
Incrementally removing the most significant events from the hot spot reveals that the excess persists, which demonstrates that the hot spot is not dominated by one or a few single events but is the result of an accumulation of neutrinos from this particular direction\cite{note:Suppl-Mat}. 
The neutrino events contributing to the excess from NGC\,1068 have been visually inspected, and show well-reconstructed horizontal $\sim$TeV energy track events with no sign of unexpected contamination or anomalies\cite{note:Suppl-Mat}. Out of the 20 events contributing most to the results presented here, 19 were included in the previous analysis \cite{Aartsen:2019fau}. Although the hot spot is thus dominated by the same neutrinos, the new data reprocessing and improved modeling of the likelihood function slightly change the precise evaluation of the events' characteristics and therefore their contribution to the likelihood function. These small but consistent recalibrations and the effects of the refined reconstructions align the events more precisely with the direction of NGC\,1068, thus significantly strengthening the neutrino association\cite{note:Suppl-Mat}.

\section*{NGC\,1068 as a neutrino source}
\begin{figure}[h]
    \centering
    \includegraphics[width=1.0\textwidth]{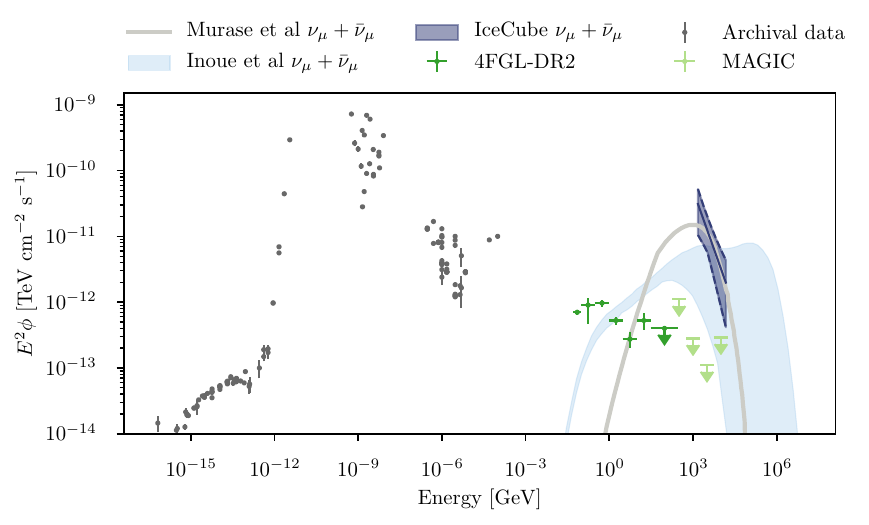}
    \caption[Spectral energy distribution of NGC\,1068]{{\bf Spectral energy distribution of NGC\,1068}. Gray points show publicly available multi-frequency measurements \cite{Chang:2019cdu}. Dark and light green error bars refer to gamma-ray measurements from \textit{Fermi}-LAT \cite{Ballet:2020hze, Fermi-LAT:2019yla} and MAGIC\cite{Acciari:2019raw}, respectively. 
    The solid, dark blue line shows the best-fit neutrino spectrum, and the corresponding blue band covers all powerlaw neutrino fluxes that are consistent with the data at $95\%$\,C.L. It is shown in the energy range between 1.5\,TeV and 15\,TeV where the flux measurement is well constrained. Two theoretical AGN core models are shown for comparison: The light blue shaded region and the gray line show the NGC\,1068 neutrino emission models from \cite{Inoue:2019yfs} and \cite{Murase:2019vdl}, respectively. Additional details on the model construction of the light blue shaded region can be found in \cite{Inoue2019abc}.}
    \label{fig:sed}
\end{figure}

The signal identified during the 3186 days period of IceCube data taking in the complete configuration consists of an estimated 79$^{+22}_{-20}$ muon-neutrino events that are present in addition to the expected background.
Scaling the 1\,TeV muon neutrino flux normalization $\Phi_{\nu_{\mu}+\bar{\nu}_{\mu}}^{1\,\textrm{TeV}}$\,=\,$(5.0\pm1.5_{stat})\times10^{-11}\,\si{TeV^{-1}.cm^{-2}.s^{-1}}$ by a factor of 3 we obtain the \textit{all-flavor} neutrino flux, assuming equal contributions from all flavors, e.g. from pion decay dominated sources and neutrino oscillations over cosmic distances \cite{Bustamante:2015waa, Arguelles:2015dca}. In the following we adopt a distance of 14.4 Mpc for NGC\,1068 \cite{Bland-Hawthorn1997}, as most commonly used in the literature, but note that there is some uncertainty with values ranging between $10.3\pm 3\,\si{Mpc}$ (from the NASA/IPAC Extragalactic Database) using the Tully-Fisher relation  and 16.5\,Mpc \cite{2020A&A...634A...1G}. The resulting redshift-corrected isotropic equivalent neutrino luminosity in the neutrino energy range from 1.5\,TeV to 15\,TeV is $L_\nu$ = $ (2.9\pm  1.1_{stat}) \times 10^{42}\,\si{erg.s^{-1}}$. This is significantly higher than the isotropic equivalent gamma-ray luminosity observed by {\it Fermi}-LAT of $1.6 \times 10^{41}\,\si{erg.s^{-1}}$ in the energy range between 100\,MeV and 100\,GeV \cite{Fermi-LAT:2019pir}, and higher than the upper limits recently reported by the MAGIC collaboration \cite{Acciari:2019raw} (see Fig.\,\ref{fig:sed}).

High-energy neutrinos are generated in or near astronomical sources as decay products of charged mesons produced in proton-proton interactions \cite{Kelner:2006tc}, or interactions between protons and low energy ambient radiation \cite{Kelner:2008ke} (for a review see \cite{Ahlers:2018fkn}). Along with those neutrinos, gamma rays are produced in the same processes through the decay of neutral mesons. After production, the neutrinos escape the site without further interactions. The photons, on the other hand, are prone to reactions depending on the optical depth of the environment. Hence the connection between neutrinos and gamma rays is highly dependent on the astrophysical environment in which the accelerated cosmic particles interact.  
NGC\,1068 \cite{Seyfert:1943} is one of the closest and best-studied Seyfert II galaxies. It hosts a Compton-thick AGN \cite{Ricci:2015rzl, Marinucci:2015fqo, Zaino:2020elj}, is also known for vigorous starburst activity \cite{Romeo:2016hms, Eichmann:2015ama} and outflows \cite{1990ApJ...355...70C}, and has therefore been discussed as a potential source of high-energy neutrinos\cite{Shapiro:1979a, Yoast-Hull:2013qfa, Lamastra:2016axo}. Recent neutrino emission models\cite{Inoue:2019yfs, Murase:2019vdl} renewed the interest in the production of neutrinos and gamma-rays within the heavily obscured environment around its core \cite{Shapiro:1979a, Shapiro:1983a}. In those models the supermassive black hole in the galaxy center provides the acceleration conditions, and the intrinsic X-ray photons generated through photon Comptonization from the accretion disk in the hot plasma above the disk, the so-called corona, provide the conditions for the production of neutrinos and at the same time also for the strong absorption of gamma rays.
\begin{figure}[h]
    \centering
    \includegraphics{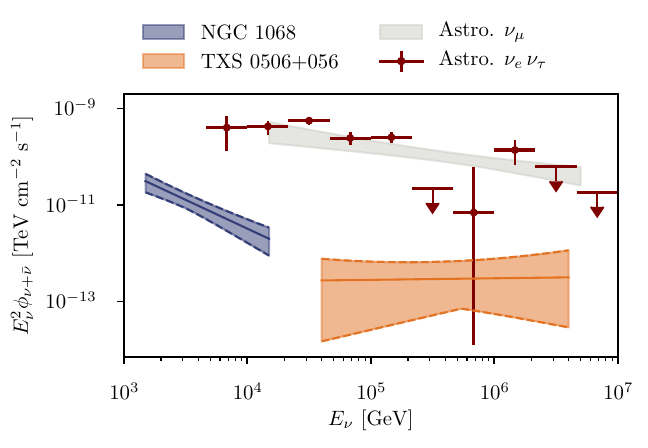}
    \caption{{\bf Comparison of point-source fluxes for NGC\,1068 and TXS\,0506+056 from this analysis with the total diffuse astrophysical neutrino flux\cite{Aartsen:2020aqd, Aartsen:2016xlq}}. Fluxes are given for a single flavor of neutrinos and anti-neutrinos assuming equal flavour ratio. The bands provide simultaneous coverage at 68\% C.L.}
    \label{fig:contribution_to_diffuse}
\end{figure}
A comparison of the neutrino flux measurement and two recent model estimations \cite{Inoue:2019yfs, Murase:2019vdl} are reported in Fig.\,\ref{fig:sed}. Since NGC\,1068 appears as a point-like neutrino source to IceCube, we cannot pinpoint directly which of its components is responsible for the neutrino emission. 

The evidence of neutrino emission from NGC\,1068 suggests that AGN could make a substantial contribution to the diffuse neutrino flux observed by IceCube.
However, Fig.\,\ref{fig:contribution_to_diffuse} demonstrates that in their respective energy ranges the sources NGC\,1068 and the blazar TXS\,0506+056 contribute no more than $\sim 1\%$ to the overall diffuse flux of astrophysical neutrinos. A more quantitative comparison is challenging: for NGC\,1068 the estimated flux ranges to energies not well covered by the measurements of the total diffuse flux, and, with a best-fit spectral index of $\hat{\gamma}=3.2^{+0.2}_{-0.2}$, the spectrum of NGC\,1068 appears to be softer than the diffuse flux with $\gamma = 2.53^{+0.07}_{-0.07}$ \cite{Aartsen:2020aqd}.
Nonetheless, it is evident that there is room for both additional bright and nearby sources, like NGC\,1068, as well as a large population of faint sources. For example, the nearby Seyfert I galaxy NGC 4151 has been suggested as one plausible candidate \cite{Inoue2019abc, PhysRevLett.66.2697, 1993ASIC..394...53S}. 
Depending on the luminosity function and cosmological evolution of the underlying population of sources, the contribution from many faint sources at large distances ($z\gtrsim 1$) to the diffuse neutrino flux can be large \cite{ Murase:2016gly, Capel:2020txc}. 

The presented evidence of neutrinos from  NGC\,1068 is insufficient to further characterize the underlying source population, for example its luminosity function and cosmological evolution. However, the observation provides some limited information on the density $\rho$ of objects with similar or greater neutrino luminosity in the local universe -- independent of the precise emission processes. Considering one source within a spherical volume of radius 14.4 Mpc \cite{Bland-Hawthorn1997}, the distance to NGC\,1068, would lead to $\rho \approx 10^{-4}\,\si{Mpc^{-3}}$.
Given that this density estimate relies on the observation of a single source, which may or may not be representative of the population, its uncertainty towards smaller densities is expected to be large and is difficult to quantify. Simulations of populations with various luminosity functions \cite{Tung:2021npo} show that it can reach well beyond an order of magnitude.
Nevertheless, despite the intrinsic uncertainty in estimating the density, the nearness of NGC\,1068 compared to TXS\,0506+056, which is approximately 100 times farther away, indicates that there are at least two populations of neutrino sources that differ in both density and luminosity by orders of magnitude.

Thanks to new reprocessing, recalibration and reconstruction of the IceCube data, we have presented here experimental evidence of neutrino emission from the active
galaxy NGC\,1068 at the 4.2$\,\sigma$ level of significance, which can be interpreted as a smoking gun signature of hadronic particle acceleration at this source. The evidence is supported by phenomenological scenarios\cite{Inoue:2019yfs, Murase:2019vdl} 
that also predict a significant absorption of gamma-ray photons. Including the blazar TXS\,0506+056 \cite{IceCube:2018cha}, 
the active galaxy NGC\,1068 is the next compelling source of high-energy neutrinos, albeit with possibly very different emission mechanisms, pointing to a population of neutrino sources with obscured gamma-ray emission.

\bibliography{scibib}

\bibliographystyle{Science}

\section*{Supplementary Materials}
Supplemental materials are freely available through the Science web portal.\\
\url{https://www.science.org}\\
\url{https://www.science.org/doi/10.1126/science.abg3395}\\
\url{https://www.science.org/doi/suppl/10.1126/science.abg3395/suppl_file/science.abg3395_sm.pdf}\\
Supplemental materials include: 
\begin{itemize}
\setlength\itemsep{0.5em}
\item{Materials and Methods}
\item{Supplementary Text}
\item{Figs. S1 to S30}
\item{Tables S1 to S4}.
\end{itemize}

\subsection*{Acknowledgments}
We dedicate this work to our recently deceased colleagues, P. Buford Price and Thomas K. Gaisser. The IceCube collaboration acknowledges significant contributions to this manuscript from Hans Niederhausen and Theo Glauch.\\

\noindent
{\bf Funding:}
The authors gratefully acknowledge the support from the following agencies and institutions:
USA {\textendash} U.S. National Science Foundation-Office of Polar Programs,
U.S. National Science Foundation-Physics Division,
U.S. National Science Foundation-EPSCoR,
Wisconsin Alumni Research Foundation,
Center for High Throughput Computing (CHTC) at the University of Wisconsin{\textendash}Madison,
Open Science Grid (OSG),
Extreme Science and Engineering Discovery Environment (XSEDE),
Frontera computing project at the Texas Advanced Computing Center,
U.S. Department of Energy-National Energy Research Scientific Computing Center,
Particle astrophysics research computing center at the University of Maryland,
Institute for Cyber-Enabled Research at Michigan State University,
and Astroparticle physics computational facility at Marquette University;
Belgium {\textendash} Funds for Scientific Research (FRS-FNRS and FWO),
FWO Odysseus and Big Science programmes,
and Belgian Federal Science Policy Office (Belspo);
Germany {\textendash} Bundesministerium f{\"u}r Bildung und Forschung (BMBF),
Deutsche Forschungsgemeinschaft (DFG),
Helmholtz Alliance for Astroparticle Physics (HAP),
Initiative and Networking Fund of the Helmholtz Association,
Deutsches Elektronen Synchrotron (DESY),
and High Performance Computing cluster of the RWTH Aachen;
Sweden {\textendash} Swedish Research Council,
Swedish Polar Research Secretariat,
Swedish National Infrastructure for Computing (SNIC),
and Knut and Alice Wallenberg Foundation;
Australia {\textendash} Australian Research Council;
Canada {\textendash} Natural Sciences and Engineering Research Council of Canada,
Calcul Qu{\'e}bec, Compute Ontario, Canada Foundation for Innovation, WestGrid, and Compute Canada;
Denmark {\textendash} Villum Fonden and Carlsberg Foundation;
New Zealand {\textendash} Marsden Fund;
Japan {\textendash} Japan Society for Promotion of Science (JSPS)
and Institute for Global Prominent Research (IGPR) of Chiba University;
Korea {\textendash} National Research Foundation of Korea (NRF);
Switzerland {\textendash} Swiss National Science Foundation (SNSF);
United Kingdom {\textendash} Department of Physics, University of Oxford.\\

\noindent
{\bf Author contributions:}
The IceCube Collaboration designed, constructed and now operates the IceCube Neutrino Observatory. Data processing and calibration, Monte Carlo simulations of the detector and of theoretical models, and data analyses were performed by a large number of collaboration members, who also discussed and approved the scientific results presented here. The manuscript was reviewed by the entire collaboration before publication, and all authors approved the final version.\\

\noindent
{\bf Competing interests:} There are no competing interests to declare.\\

\noindent
{\bf Data and materials availability:}
Additional data and resources are available from the IceCube data archive at \url{https://icecube.wisc.edu/science/data-releases}. For each data sample these include the events, neutrino effective areas, background rates, and other supporting information in machine-readable formats.

\pagebreak
\subsection*{IceCube Collaboration$^{\ast}$:}
R. Abbasi$^{17}$,
M. Ackermann$^{61}$,
J. Adams$^{18}$,
J. A. Aguilar$^{12}$,
M. Ahlers$^{22}$,
M. Ahrens$^{51}$,
J.M. Alameddine$^{23}$,
C. Alispach$^{28}$,
A. A. Alves Jr.$^{31}$,
N. M. Amin$^{43}$
% ...
\\
\\
% ...
$^{12}$ Science Faculty CP230, Universit{\'e} Libre de Bruxelles, B-1050 Brussels, Belgium \\
% ...
$^{17}$ Department of Physics, Loyola University Chicago, Chicago, IL 60660, USA \\
$^{18}$ Department of Physics and Astronomy, University of Canterbury, Private Bag 4800, Christchurch, New Zealand \\
% ...
$^{22}$ Niels Bohr Institute, University of Copenhagen, DK-2100 Copenhagen, Denmark \\
$^{23}$ Department of Physics, Technical University Dortmund, D-44221 Dortmund, Germany \\
% ...
$^{28}$ D{\'e}partement de physique nucl{\'e}aire et corpusculaire, Universit{\'e} de Gen{\`e}ve, CH-1211 Gen{\`e}ve, Switzerland \\
% ...
$^{31}$ Karlsruhe Institute of Technology, Institute for Astroparticle Physics, D-76021 Karlsruhe, Germany  \\
% ...
$^{43}$ Department of Physics and Astronomy,  University of Delaware, Newark, DE 19716, USA \\
% ...
$^{51}$ Oskar Klein Centre and Department of Physics, Stockholm University, SE-10691 Stockholm, Sweden \\
% ...
$^{61}$ Deutsches Elektronen-Synchrotron, D-15738 Zeuthen, Germany \\
% ...

\noindent $^\ast$E-mail: analysis@icecube.wisc.edu

\end{document}